%% file: paper.tex
\PassOptionsToPackage{dvipsnames}{xcolor}
\documentclass[10pt,sigconf,nonacm,screen]{acmart}

\usepackage[all]{nowidow}
\usepackage{xcolor}
\usepackage{subcaption}
\usepackage{hyperref}
\usepackage{acronym}

\usepackage[capitalise,noabbrev]{cleveref}
\crefformat{section}{\S#2#1#3} 
\crefformat{subsection}{\S#2#1#3}
\crefformat{subsubsection}{\S#2#1#3}
\crefrangeformat{section}{\S#3#1#4 to~\S#5#2#6}
\crefrangeformat{subsection}{\S#3#1#4 to~\S#5#2#6}
\crefrangeformat{subsubsection}{\S#3#1#4 to~\S#5#2#6}

\newcommand{\name}{\textsc{GeoFF}}

\begin{document}

\author{Natalie Carl}
\affiliation{
    \institution{TU Berlin \& ECDF}
    \city{Berlin}
    \country{Germany}}
\email{nc@3s.tu-berlin.de}
\orcid{0009-0000-5991-9255}

\author{Trever Schirmer}
\affiliation{
    \institution{TU Berlin \& ECDF}
    \city{Berlin}
    \country{Germany}}
\email{ts@3s.tu-berlin.de}
\orcid{0000-0001-9277-3032}

\author{Tobias Pfandzelter}
\affiliation{
    \institution{TU Berlin \& ECDF}
    \city{Berlin}
    \country{Germany}}
\email{tp@3s.tu-berlin.de}
\orcid{0000-0002-7868-8613}

\author{David Bermbach}
\affiliation{
    \institution{TU Berlin \& ECDF}
    \city{Berlin}
    \country{Germany}}
\email{db@3s.tu-berlin.de}
\orcid{0000-0002-7524-3256}

\title{\name{}: Federated Serverless Workflows with Data Pre-Fetching}


\begin{abstract}
    Function-as-a-Service (FaaS) is a popular cloud computing model in which applications are implemented as workflows of multiple independent functions.
    While cloud providers usually offer composition services for such workflows, they do not support cross-platform workflows forcing developers to hard-code the composition logic.
    Furthermore, FaaS workflows tend to be slow due to cascading cold starts, inter-function latency, and data download latency on the critical path.
    
    In this paper, we propose \name{}, a serverless choreography middleware that executes FaaS workflows across different public and private FaaS platforms, including ad-hoc workflow recomposition.
    Furthermore, \name{} supports function pre-warming and data pre-fetching.
    This minimizes end-to-end workflow latency by taking cold starts and data download latency off the critical path.
    In experiments with our proof-of-concept prototype and a realistic application, we were able to reduce end-to-end latency by more than 50\%.
\end{abstract}

\maketitle

\input{sections/1_intro}
\input{sections/2_background.tex}
\input{sections/3_approach.tex}
\input{sections/4_evaluation.tex}
\input{sections/5_discussion.tex}
\input{sections/6_conclusion.tex}

\balance

\bibliographystyle{ACM-Reference-Format}
\bibliography{bibliography}

\end{document}

%% file: sections/1_intro.tex
\section{Introduction}\label{sec:introduction}

Serverless computing in the form of Function-as-a-Service (FaaS) is a service offering in cloud computing that allows developers to focus on writing their applications in small, stateless functions~\cite{paper_bermbach2021_cloud_engineering}.
Management of operational concerns, such as infrastructure and the execution environment, are handled by the serverless platform provider~\cite{Wang_2018_Peeking}.
In turn, FaaS makes it easier for developers to build applications organized as workflows of multiple functions.

Options for FaaS platforms span the entire edge-to-cloud continuum, catering to various deployment needs from commercial options, including Amazon,\!\!\footnote{\url{https://aws.amazon.com/lambda/}} Google,\!\!\footnote{\url{https://cloud.google.com/functions}} and Microsoft,\!\!\footnote{\url{https://azure.microsoft.com/products/functions}} to open-source platforms that enable FaaS deployments in the private and hybrid cloud, e.g.,~\cite{OpenFaaS,Nuclio,OpenWhisk}.
FaaS platforms such as tinyFaaS~\cite{paper_pfandzelter2020_tinyfaas}, NanoLambda~\cite{NanoLambda}, ServerLedge~\cite{russo_2023_serverledge}, or Lean OpenWhisk~\cite{LeanOpenWhisk} extend the FaaS model to the edge.

These offerings, however, lack integration due to their use of different control mechanisms and communication formats, which limits their ability to communicate with each other.
In use cases that require using multiple serverless providers, users need to distribute parts of their application across different compute locations and serverless offerings.
This requires the ability to execute functions in different environments and to execute workflows across providers using orchestration or choreography.
At the moment, most FaaS platforms come with orchestrator support -- to our knowledge, however, none of these support cross-provider composition.
As a result, developers currently have to hard-code cross-provider workflows.

As a second problem, serverless computing is often used for batch tasks and data analysis jobs~\cite{Eismann_2021_State,Werner_2024_Reference,muller2020lambada,splitserve}.
Once a function in such a workflow has been called, it will often spend significant time downloading additional input data from, e.g., a cloud datastore~\cite{hellerstein2018serverless}.
This, combined with cascading cold starts in long workflows~\cite{paper_bermbach2020_faas_coldstarts,daw_xanadu_2020} leads to unnecessarily high end-to-end workflow latency and, thus, also increases cost.

Here, we see opportunities to improve FaaS workflow efficiency:
First, workflows with external data dependencies can benefit from function shipping, where functions are executed closer to the data they depend on to reduce transfer time, cost, and data transfer across cloud boundaries. 
Second, workflow knowledge can be used to further reduce wait times: Through overlaying communication and computation, a workflow step can load data during the previous step's execution, which further shortens workflow durations.
To the best of our knowledge, these optimizations have not yet been holistically proposed and evaluated for federated cloud and edge-to-cloud FaaS workflows.

In this paper, we propose \name{}, a choreography middleware that allows (1) serverless workflows to comprise functions running on different cloud providers, private data centers, and edge devices in different regions, (2) dynamically changing workflow choreographies across those boundaries, and (3) automatic function pre-warming and data pre-fetching to mitigate cold starts, reduce the effect of latency introduced by loading data, and to minimize total workflow durations.
We make the following contributions:
\begin{itemize}
    \item We propose \name{}, a system that optimizes workflow performance in federated serverless applications through data pre-fetching and function shipping (\cref{sec:approach}).
    \item We implement a prototype of \name{} for tinyFaaS, AWS Lambda, and Google Cloud Functions (\cref{subsec:impl}).
    \item We evaluate \name{} with multiple use cases in public, hybrid, and private cloud settings (\cref{sec:evaluation}).
\end{itemize}

%% file: sections/2_background.tex
\section{Background \& Related Work}\label{sec:background}

In FaaS workflows, functions can be composed into function chains, where the output of one function serves as the input of the following function to construct large and complex serverless applications.
These workflows are commonly represented as directed acyclic graphs and can include different relationships between functions, such as fan-in and fan-out patterns~\cite{daw_xanadu_2020}.
Commercial solutions for FaaS workflow management include AWS Step Functions\footnote{\url{https://aws.amazon.com/step-functions/}} and Google Cloud Workflows,\!\!\footnote{\url{https://cloud.google.com/workflows}} which manage workflows centrally, i.e., a central orchestrator invokes the first function, collects its result, invokes the next function and so forth~\cite{paper_bermbach2020_faas_coldstarts}.

This approach will often lead to cascading cold starts~\cite{paper_bermbach2020_faas_coldstarts,daw_xanadu_2020} and cause costs for data movement and control overhead in the orchestrator.
`Cold start cascades' can be remedied through function pre-warming~\cite{paper_bermbach2020_faas_coldstarts}, i.e., using the available workflow knowledge to trigger cold starts off the critical path, a strategy also adapted in Xanadu~\cite{daw_xanadu_2020}.

To reduce data transfer costs, Tang and Yang propose \emph{Lambdata}~\cite{tang_lambdata_2020}, which caches function outputs locally in addition to uploading it to cloud storage.
By scheduling subsequent workflows steps on the same worker instance, additional latency from uploading and downloading data can be avoided.
However, this approach requires modifications of the underlying platform, which is infeasible in a commercial platform, let alone in a federated setting across cloud boundaries.

To reduce data movement between functions in a workflow and a central orchestrator, function workflows can also be designed in a decentralized manner, where each function directly calls its successor.
This allows optimizations such as \emph{SAND}~\cite{akkus_sand_2018}, where functions of the same applications are grouped into a container to reduce the amount of cold starts within a workflow and the communication latency between functions.
Similarly, \emph{SPRIGHT}~\cite{10.1145/3544216.3544259} uses eBPF-based networking to improve the communication latency from one function to the next, further reducing the execution duration of a FaaS workflow.
In previous work~\cite{paper_schirmer2022_fusionize,paper_schirmer2023_fusionizepp}, we have shown that merging workflow tasks into the same cloud function --- akin to compiler inlining --- can also reduce end-to-end latency as it avoids FaaS platform overheads.

Beyond deploying FaaS functions on a single cloud platform, it can be desirable to run workflows across different public and private cloud platforms:
For example, there might be services that are exclusive to one provider, legal requirements to process sensitive data in certain regions, more options for optimized function placement, or existing systems that are to be integrated.
If there are data dependencies in a specific cloud or edge location, shipping functions to that data can improve total workflow durations and be cheaper than shipping data, especially considering cloud egress costs~\cite{AWSTransferCost}.
A challenge arises with workflow portability and interoperability across different cloud providers and regions.
FaaS functions are structured differently depending on the platform they are deployed on, which complicates federation efforts to deploy, execute, and move functions seamlessly across multiple providers and regions.

\emph{XFaaS}~\cite{Khochare_2023_xfaas} enables cross-platform FaaS workflows through a central orchestrator.
It optimizes workflow execution time through function fusion and function shipping.
Chard et al.~\cite{chard_2020_funcX,babuji_2021_faas_science} propose \emph{funcX}, a federated FaaS-platform for scientific computations. 
The platform takes heterogeneous and distributed resources into account, which enables the deployment and invocation of functions across the cloud--edge continuum.
These approaches are excellent ways to improve the developer experience of FaaS workflows; yet, an implementation relying on a central orchestration component incurs performance overheads, e.g., with data and control movement.

There are multiple approaches focusing on FaaS in fog and edge environments: 
Cheng et al.~\cite{cheng_2019_fog_function} introduce a programming model to orchestrate serverless fog computing, focusing on data-intensive applications.
Instead of using traditional events, the orchestration mechanism uses the availability of raw and processed input data, the nodes' available resources, and user defined usage contexts to manage workflows.
\emph{Serverledge}~\cite{russo_2023_serverledge} is a distributed FaaS-platform consisting of multiple cloud- and edge-nodes.
Requests can be sent to and invoked at every node, and the platform supports horizontal (to other edge nodes) and vertical (to the cloud) offloading.
Serverledge is a complete FaaS platform that is distributed but not federated by design.
Likewise, AuctionWhisk~\cite{paper_bermbach2021_auctionwhisk,paper_bermbach2020_auctions4function_placement} can distribute function execution over different FaaS nodes on the edge-to-cloud continuum.

Pre-fetching extends pre-warming by downloading required data from storage services before the function invocation to mitigate latency overheads.
Through pre-fetching, different workflow steps overlap, which optimizes the total workflow duration compared to sequential workflow execution.
Tang and Yang~\cite{tang_lambdata_2020} argue that this does not change the FaaS workflow's execution cost, yet there are benefits in execution duration when moving data transfer outside the workflow's execution path.
Pre-fetching can also be combined with caching data in function instances, as implemented, e.g., in \emph{Faa\$t}~\cite{Romero_2021_faast}.

Overall, we are not aware of any approach supporting cross-provider choreographies with client-initiated ad-hoc recomposition that uses workflow knowledge to pre-warm functions, to pre-fetch data, and to ship functions to data if needed.

%% file: sections/3_approach.tex
\section{Cross-Provider Choreographies with \name{}}\label{sec:approach}

\begin{figure}
    \centering
    \includegraphics[width=\linewidth]{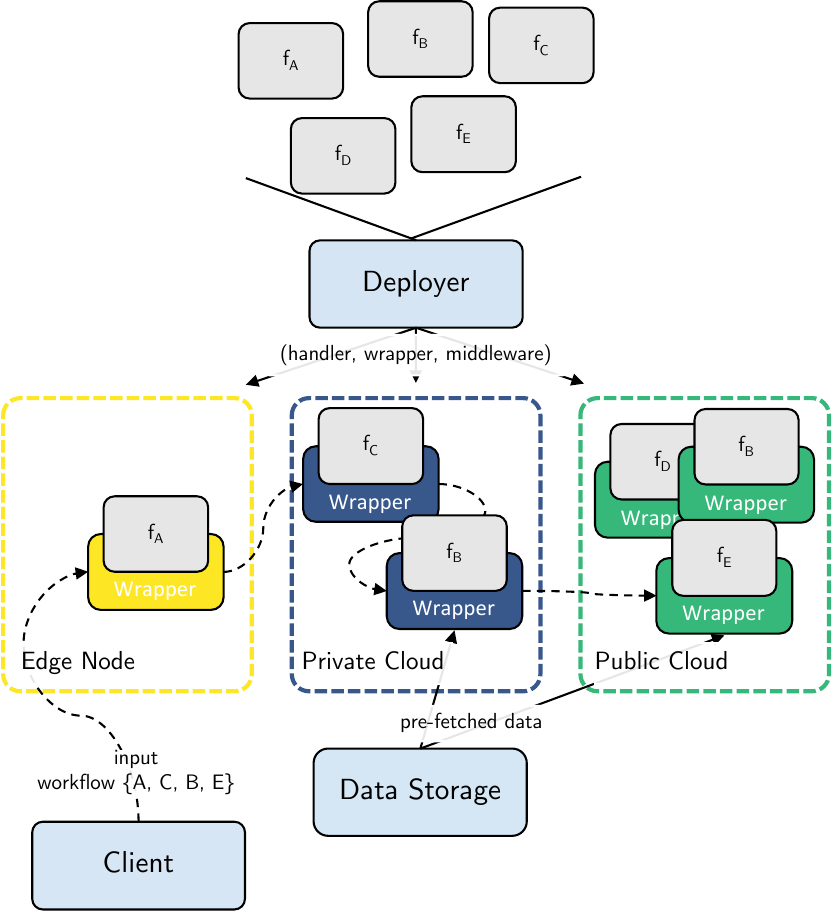}
    \caption{
        The \name{} deployer takes code of functions (here $f_A$ to $f_E$), dependencies, and a deployment configuration to deploy an application across different public clouds, private clouds, and edge nodes. Each resulting FaaS function consists of the developer's function handler, a platform-specific wrapper, and the choreography middleware that handles workflow execution and data pre-fetching. Clients start workflows by invoking the first step with a function input and a workflow specification.
    }
    \label{fig:architecture}
\end{figure}

\name{} improves workflow execution across cloud and edge FaaS platform.
Specifically, \name{} includes a middleware that (i) allows writing a FaaS function once and deploying it to any FaaS platform, (ii) includes a co-deployed, decentralized choreography middleware in each function instance that can parse dynamic workflow specifications to execute workflows without a central orchestrator, and (iii) lets functions specify external data dependencies that the middleware can pre-fetch before the function is called in the workflow.
\cref{fig:architecture} provides an overview of the architecture of \name{}.

\subsection{Federated Deployment}

To create an application, a developer writes a deployment specification, which determines which function should be deployed on which platforms in which regions.
Together with the function code and a list of dependencies, the \emph{deployer} automatically deploys all functions according to the specification.
The functions are written in a platform-independent manner.
This allows us to move functions to different and multiple platforms without adjusting their code.
In addition to the function handler, the deployer packages a choreography middleware (cf.~\cref{subsec:workflows}) and a platform-specific wrapper into every deployed FaaS function.
Each of the developer's function handlers takes the function inputs and, if applicable, a path to pre-fetched data as input, as the data will be automatically downloaded by the wrapper and middleware during the workflow execution.

\subsection{Workflow Choreographies}\label{subsec:workflows}

In order for \name{} to effectively manage geo-distributed workflows, we avoid relying on a central component for orchestration to prevent a potential performance bottleneck.
In addition, this would increase the total duration of workflows because of the need to send additional messages to and from the orchestrator.
This becomes particularly problematic when the orchestrator is located far away from the workflow steps in the network.

To this end, \name{} includes a choreography middleware that is co-deployed with every function handler.
This middleware serves two purposes:
First, it handles the execution of workflows by invoking other functions and keeping track of the workflow state.
Second, it downloads data that is to be processed by the function handler (cf.~\cref{subsec:prefetch}).

To invoke a workflow, a user sends a request to the first step with that function's input and a specification of the workflow, which will be used by the choreography middleware to execute the entire workflow.
This workflow specification includes information about which workflow steps to invoke in which order and determines which data will be pre-fetched by the middleware.
To support ad-hoc recomposition, the specification only needs to be available at runtime, not during deployment.
Consequently, a workflow can be rerouted on a per-request basis, and no re-deployment is necessary to take different routes.
Further, the same function can be deployed to different platforms simultaneously, which increases the application's fault tolerance and decreases the dependence on a specific provider.
As part of the call, the workflow specification controls on which provider the function is actually invoked.
While this adds complexity for clients, we would expect that such a workflow specification will in practice be handled by a client-side library.

\subsection{Data Pre-Fetching}\label{subsec:prefetch}

\begin{figure}
    \centering
    \includegraphics[width=\linewidth]{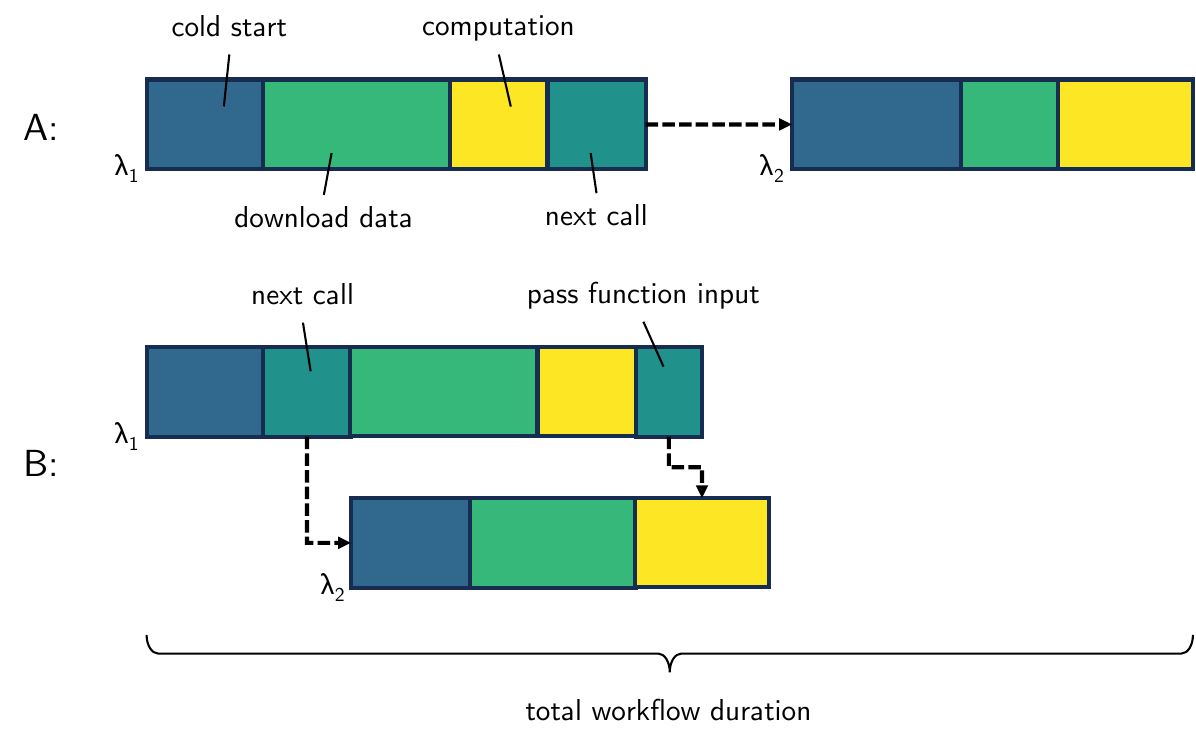}
    \caption{
        Workflow A is executed sequentially: The second step $\lambda_2$ is invoked only after completion of $\lambda_1$.
        For Workflow B, in contrast, $\lambda_2$ already experiences its cold start and downloads the required data while $\lambda_1$ is still executing its function logic.
        This reduces the total workflow duration.
    }
    \label{fig:prefetching}
\end{figure}

\cref{fig:prefetching} shows two workflows, each consisting of two functions, and the generic parts a workflow step typically performs.
First, if there is no function instance available, a new instance will be created, resulting in additional cold start latency.
Next, a function has to download the data it processes in the current invocation.
Once the download is finished, the function logic is executed to process the data.
Lastly, the workflow's next step is invoked.

To execute a workflow correctly, the order of execution of the workflow steps is important.
However, not every part of a function's execution requires the previous step to be finished.
In the second step, only the execution of the function handler (see the yellow boxes in \cref{fig:prefetching}) depends on the output of the previous function handler.
In contrast, the cold start and downloading data do not depend on the first step.

In \name{}, we use this characteristic to optimize workflow duration by moving the invocation of the second workflow step forward.
In the optimum, we move the second step such that the computation phases directly follow each other but do not overlap, as can be seen in workflow \texttt{B} in \cref{fig:prefetching}.
This way, the total execution duration does not change, but the total workflow duration will be reduced because we effectively subtract the amount of time it takes to download data form the total workflow duration through overlapping multiple workflow steps.
Consequently, the two workflow steps have to communicate twice: First, the initial step pokes the following step without passing any arguments.
Later, after the first function has finished processing its data, it sends the now determined function inputs to the following step.

A challenge that arises is predicting how much sooner to start following workflow steps.
If not timed perfectly, one of the steps will have to wait, resulting in unnecessary double billing~\cite{Baldini_2017_Trilemma}.
We believe that such timing could easily be learned from monitoring data, so the call time would gradually converge towards the optimal call time after several invocations.
This, however, is beyond the scope of this work.

%% file: sections/4_evaluation.tex
\section{Evaluation}
\label{sec:evaluation}

To evaluate \name{}, we implement a proof-of-concept prototype (\cref{subsec:impl}).
With three experiments, we analyze the impact of data pre-fetching (\cref{subsec:prefetching}) and function shipping in FaaS workflows (\cref{subsec:funcshipping}) before evaluating the impact of native pre-fetching (i.e., directly integrating \name{} into the FaaS platform, \cref{subsec:native}).
We make our prototype and experiment results available as open-source\footnote{\url{https://github.com/valentin-carl/federation}}\!\!.

\subsection{Proof-of-Concept Prototype}
\label{subsec:impl}
Our prototype of \name{} consists of three components: the deployer, platform-specific wrappers, and the choreography middleware.

The deployer uses the Serverless Framework\footnote{\url{https://www.serverless.com/}} to deploy functions on different FaaS platforms.
Since the Serverless Framework only includes support for AWS Lambda and Google Cloud Functions, we add a plugin that enables the Serverless Framework to deploy functions to tinyFaaS~\cite{paper_pfandzelter2020_tinyfaas}.
To move functions between platforms without adjusting their code, we co-deploy platform-specific wrappers that serve as entry points for the user-specified functions.
To create a deployment, the user has to provide the function code and deployment specification, which can be reconfigured dynamically.
In preliminary experiments, we found the wrapper to introduce a call overhead of less than 1ms, which is negligible for most FaaS use cases~\cite{Eismann_2021_State} -- especially compared to the latency savings from data pre-fetching and function pre-warming.

The choreography middleware that is co-deployed in every function enables the execution of serverless workflows across cloud platforms by abstracting the platform specific interface.
It handles the invocation of workflow steps, pre-fetching of data, and passing arguments between functions.

This allows the routing between workflow steps to be determined at the time of workflow invocation and instead of at deploy-time.
As the workflow for a specific invocation is passed with every request, \name{} can be reconfigured for every invocation.

To implement pre-fetching in AWS Lambda and Google Cloud Functions, the previous workflow step communicates with its successors twice: first to initialize the pre-fetch (before the actual workflow step finishes), and a second time to pass the result of its own computation (cf.~\cref{fig:prefetching}).
Because public clouds disable outside traffic from reaching FaaS functions, the \emph{non-native} pre-fetching uses the cloud-based object storage AWS S3 to buffer inputs between workflow steps.
Because the actual function invocation happens during the pre-fetching phase and the function input is passed via S3, the public cloud implementations do not support synchronous calls.
In the case of tinyFaaS, having provider-side control over the platform allows us to implement \emph{native pre-fetching}: The platform intercepts the pre-fetching call and loads the data without calling the function code itself.
As a result, the function code is only called when the output of the previous step has arrived, thus, supporting synchronous invocations.

We implement the prototype and use cases in Python.

\subsection{Experiment: Data Pre-Fetching}
\label{subsec:prefetching}

\begin{figure}
    \centering
    \includegraphics[width=\linewidth]{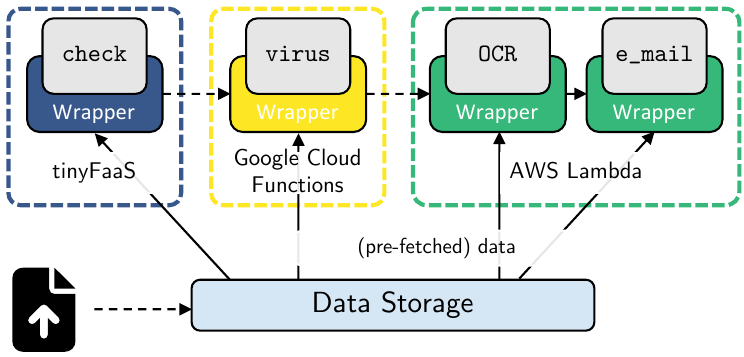}
    \caption{
        The document preparation workflow is adapted from our previous work~\cite{schirmer2023profaastinate}:
        Users initiate the pre-check to ensure the PDF's correctness and store it in an object storage system.
        Subsequently, asynchronous processes are triggered for virus scanning, optical character recognition, and e-mail notifications.
    }
    \label{fig:usecase_1}
\end{figure}

The purpose of the first experiment is to determine the impact of pre-fetching data throughout a workflow spanning multiple different cloud platforms.
As shown in \cref{fig:usecase_1}, we adapt the document processing use case from Schirmer et al.~\cite{schirmer2023profaastinate}.
This workflow consists of four functions: First, a user uploads a PDF file through the \texttt{check} function.
This function verifies that the file uploaded by the user is actually a PDF file and, if successful, uploads the document to storage.
Second, the \texttt{virus} function analyzes whether the file contains malicious content.
Third, the \texttt{OCR} function performs optical character recognition in order for the last step to be able to parse the PDF file's contents.
As the last step, the \texttt{e\_mail} function reads the file contents and sends them with a notification to some user.
Note that in this use case, all functions except for the first require downloading external data.

We deploy this use case to AWS Lambda (region \texttt{us-east-1}), Google Cloud Functions (region \texttt{europe-west10}), and tinyFaaS as an edge-node.
The tinyFaaS instance runs on an \texttt{e2-highcpu-8} Google Compute Engine instance, with 8 vCPUs and 8 GB of RAM running Ubuntu 22.04 LTS.
The VM is also deployed in the region \texttt{europe-west10}.
We use this tinyFaaS deployment configuration throughout all three experiments.

For this experiment, the \texttt{check} function is deployed to the tinyFaaS instance to be close to end users and thus save bandwidth, \texttt{virus} to Google Cloud Functions, and the last two steps are deployed to AWS Lambda.
As outlined above, one motivation for federated serverless applications is that some code might only be supported for one specific platform.
In our case, this applies to the OCR functions, as they require external dependencies that were not available as Python libraries, so we can only deploy them to AWS Lambda as custom Docker containers.
Even without this, the edge-deployed \texttt{check} function would already cause the workflow to span multiple providers.

In the experiment, we compare two workflow configurations.
In the first, all steps (except for the initial call) pre-fetch the data they require.
In the second, which we use as a baseline for comparison, no steps pre-fetch data and the entire workflow is executed completely sequentially.
For each configuration, we send one request per second for 30 minutes and measure the total workflow duration.

\begin{figure}
    \centering
    \includegraphics[width=\linewidth]{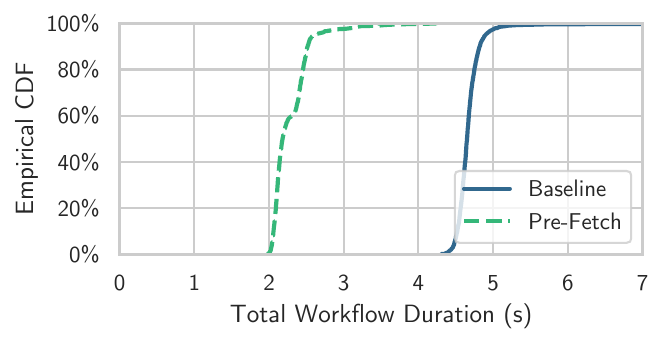}
    \caption{
        This graph shows the cumulative distribution up to the 99\textsuperscript{th} percentile of total workflow duration measurements for the document processing workflow of the data pre-fetching experiment. 
        In the median, pre-fetching reduces the total workflow duration by 53.02\% compared to the baseline case.
    }
    \label{fig:results_e1}
\end{figure}

\cref{fig:results_e1} shows the results of the first experiment.
Using pre-fetching, the median total workflow duration of the document processing use case was 2.19s compared to 4.65s in the baseline case.
In this experiment, pre-fetching improves the total workflow duration by 53.02\%.

\subsection{Experiment: Function Shipping}
\label{subsec:funcshipping}

\begin{figure}
    \centering
    \includegraphics[width=\linewidth]{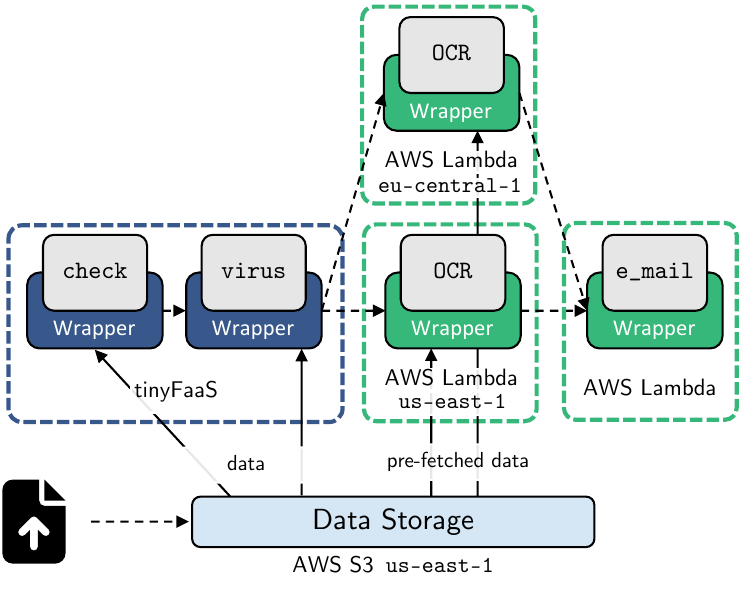}
    \caption{
        In the second use case (function shipping), only the \texttt{OCR} function pre-fetches. 
        The function is deployed to the AWS regions \texttt{eu-central-1} and \texttt{us-east-1}.
    }
    \label{fig:usecase_2}
\end{figure}

The purpose of the second experiment is to determine the effect of moving a function that requires external data from a location far away to a location that is close to where the external data is stored.
We adapt the document processing workflow so that only the \texttt{OCR} function requires downloading files.
The first and second step, the \texttt{check} and \texttt{virus} functions, are deployed to the tinyFaaS node. 
Next, the \texttt{OCR} and \texttt{e\_mail} functions are deployed on AWS Lambda.
The \texttt{e\_mail} function runs in the region \texttt{us-east-1}.
As the \texttt{OCR} function is the only part of this workflow that requires external data, we run two variations of this experiment: in the first, the \texttt{OCR} function is deployed in the region \texttt{eu-central-1}.
In the second variation, we move the function to the region \texttt{us-east-1}, where the data is also stored.
For both variations, the function uses pre-fetching to download the data from object storage.
Similarly to the first experiment, we send one request per second for 30 minutes for each variation and measure the total workflow duration.

\begin{figure}
    \centering
    \includegraphics[width=\linewidth]{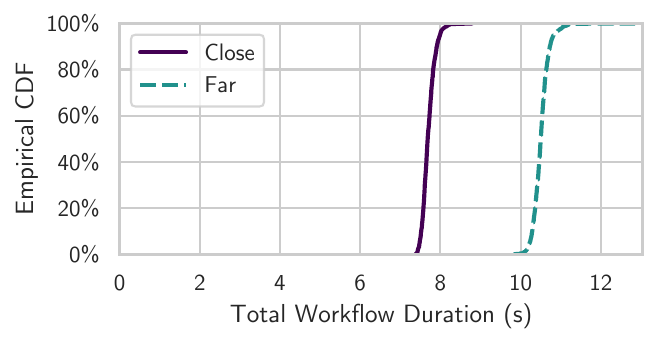}
    \caption{
        In the function shipping experiment, the \texttt{OCR} function is moved from AWS region \texttt{eu-central-1} (far) to \texttt{us-east-1} (close), where the data this function pre-fetches is stored.
        Using pre-fetching in both cases, moving the function to the data it pre-fetches improves total workflow duration by 26.90\% in the median.
    }
    \label{fig:results_e2}
\end{figure}

\cref{fig:results_e2} shows the results of the second experiment.
In the second experiment, we measure the effect of moving a function close to where the data it processes is stored.
With the data stored in the AWS region \texttt{us-east-1} and the function deployed to region \texttt{eu-central-1}, we measure a median total workflow duration of 10.47s.
Moving the function to the same region as the data reduces this median to 7.65s, a reduction of 26.90\%.

\subsection{Experiment: Native Pre-Fetching}
\label{subsec:native}

\begin{figure}
    \centering
    \includegraphics[width=\linewidth]{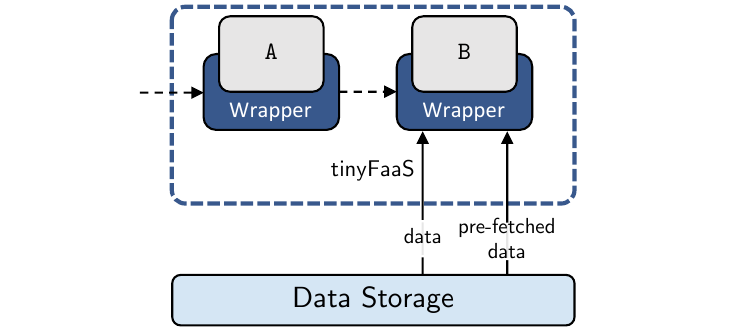}
    \caption{
        The native pre-fetching experiment compares the effect of pre-fetching on total workflow duration for two function deployed to the same edge node. 
        Function \texttt{B} loads 256 KB of data, which can be pre-fetched while function \texttt{A} is running.
    }
    \label{fig:usecase_3}
\end{figure}

The purpose of the third experiment is evaluating the effect of \emph{native} pre-fetching in one step on the total workflow duration.
The use case comprises only two functions: Function \texttt{A} performs arbitrary computations for five seconds, after which the second function is called.
Function \texttt{B} depends on a 256KB file, which is either pre-fetched with or downloaded after \texttt{A} finishes.

We deploy both functions to the tinyFaaS node where we have implemented native pre-fetching.
Again, we split the experiment into two variations: In the first variation, function \texttt{B} pre-fetches the data from S3.
In the baseline case, the data is only downloaded after function \texttt{A} has terminated.
Again, we send one request per second for 30 minutes for each variation and measure the total workflow duration.

\begin{figure}
    \centering
    \includegraphics[width=\linewidth]{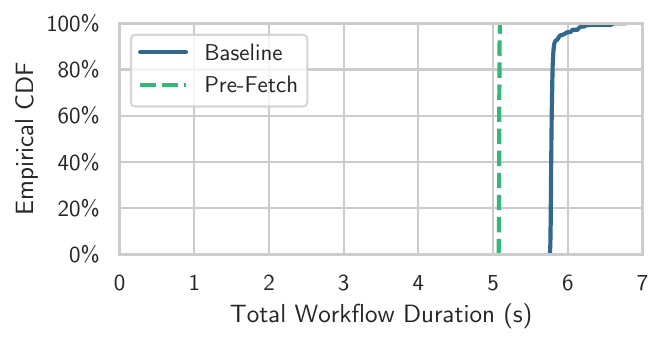}
    \caption{
        The native pre-fetching experiment compares the effect of pre-fetching on total workflow duration for a two-function use case, where both functions are located on the same edge node.
        Here, total workflow duration is reduced by 12.08\% in the median using pre-fetching.
    }
    \label{fig:results_e3}
\end{figure}

As the experiment shows (cf., \cref{fig:results_e3}), the workflows takes 5.87s in the median to execute in the baseline case and 5.08s using pre-fetching, which constitutes an improvement of 12.08\%.

%% file: sections/5_discussion.tex
\section{Discussion \& Future Work}\label{sec:discussion}

Our evaluation has shown how data pre-fetching and function shipping with \name{} can reduce total FaaS workflow duration.
Nevertheless, we see several avenues for future improvements. 

\subsection{Performance}

The biggest performance improvements with \name{} can be achieved for workflows that pre-fetch large amounts of data, e.g., data science workflows\cite{Werner_2024_Reference}.
In contrast, in smaller applications without data dependencies and functions that run on the order of milliseconds, the relative performance impact of the wrapper can increase.

\subsection{Native Pre-Fetching}

While direct platform support leads to lower overheads during pre-fetching, even our implementation on top of public cloud platforms showed reduced workflow durations and cost.
Additionally, all workflow steps executed in the public cloud have to be asynchronous. 
Synchronous requests with pre-fetching can only be implemented with direct platform support.
In future work, we hope to reduce the impact of pre-fetching in public FaaS platforms further, e.g., by directly communicating with the function instance~\cite{Copik_2023_fmi}.

\subsection{Automated Function Placement}

While \name{} has support for seamlessly moving functions between platforms, developers currently have to decide on the deployment configuration.
Automating this would enable optimized function placement during runtime, e.g., functions can be moved towards data to reduce transfer costs and applications can be seamlessly migrated to fulfill dynamically changing business requirements.
Optimization- or machine learning-based approaches such as~\cite{paper_bellmann2021_predictive_replication_markov} could help the client to decide which function instances to include in the triggered workflow instance.

\subsection{Automated Path Optimization}

Due to using a decentralized choreography middleware, the client sends the workflow specification with every invocation.
We are currently working on using this to optimize workflow specifications automatically, depending on client requirements.
This programming interface should then be used by a routing optimizer that optimizes workflow choreographies on the clients' behalf.

\subsection{Optimized Pre-Fetch Timing}

Currently, the middleware signals the next workflow step to pre-fetch external data as soon as the current step is invoked.
Starting pre-fetching too early leads to unnecessary double-billing and wasted storage resource but reduces total workflow duration.
Starting it too late leads to longer workflow durations but limits wasted resources, i.e., we currently minimize workflow duration while accepting the upper bound on double-billing costs.
The decision on how early the pre-fetching of data should be started could be automated using runtime insights, which we plan to explore in future work.

%% file: sections/6_conclusion.tex
\section{Conclusion}\label{sec:conclusion}

FaaS is a popular cloud computing model in which applications are implemented as workflows of multiple independent functions while providers handle the application lifecycle.
Even though cloud providers usually offer orchestration services for FaaS workflows, e.g., AWS Step Functions, there are currently no solutions that support cross-platform workflows, thus,  forcing developers to hard-code the composition logic into their application.
Furthermore, FaaS workflows tend to be slow due to cascading cold starts, inter-function latency, and data download latency on the critical path.

In this paper, we proposed \name{}, a serverless choreography middleware that automatically executes FaaS workflows across different public and private edge-to-cloud FaaS platforms, including ad-hoc workflow recomposition as triggered by application clients.
Furthermore, \name{} supports function pre-warming and data pre-fetching.
This minimizes end-to-end workflow latency by taking cold starts and data download latency off the critical path.
In experiments with our proof-of-concept prototype and a realistic document processing application, we were able to reduce end-to-end workflow latency by more than 50\%.